\documentclass{aa}  
\usepackage{graphicx}
\usepackage{natbib}
\bibpunct{(}{)}{;}{a}{}{,}
\begin{document}
   \title{Formation of methyl formate and other organic species in the warm-up phase of hot molecular cores}
   \authorrunning{Garrod \& Herbst}
   \titlerunning{Grain-surface formation of methyl formate}
   \subtitle{}

   \author{R. T. Garrod\inst{1}
          \and
          E. Herbst\inst{1,2}
          }

 \offprints{R. T. Garrod, \email{rgarrod@mps.ohio-state.edu}}
   \institute{Department of Physics,
    The Ohio State University, Columbus, OH 43210, USA \and
    Departments of Astronomy and Chemistry, The Ohio State University, Columbus, OH 43210,USA}
             

   \date{Received xxx 00, 0000; accepted xxx 00, 0000}

 
  \abstract
   {xxx}
   {The production of saturated organic molecules in hot cores and corinos is not well understood.  The standard approach is to assume that, as temperatures heat up during star formation, methanol and other species evaporate from grain surfaces and undergo a warm gas-phase chemistry at 100 K or greater to produce species such as methyl formate, dimethyl ether, and others.  But a series of laboratory results shows that protonated ions, typical precursors to final products in ion-molecule schemes, tend to fragment upon dissociative recombination with electrons rather than just ejecting a hydrogen atom.  Moreover, the specific proposed reaction to produce protonated methyl formate is now known not to occur at all. }
   {We utilize a gas-grain chemical network to probe the chemistry of the relatively ignored stage of hot core evolution during which the protostar switches on and the temperature of the surrounding gas and dust rises from 10 K to over 100 K. During this stage, surface chemistry involving heavy radicals becomes more important as surface hydrogen atoms tend to evaporate rather than react. }
   {Our results show that complex species such as methyl formate, formic acid, and dimethyl ether can be produced in large abundance during the protostellar switch-on phase, but that both grain-surface and gas-phase processes help to produce most species.  The longer the timescale for protostellar switch-on, the more important the surface processes.}
   {xxx}


   \maketitle
%

\section{Introduction}

Various large molecules including methyl formate (HCOOCH$_3$) have been detected in a number of hot molecular cores and corinos \cite[]{blake87a,hatchell98a,nummelin00a,cazaux03a,bottinelli04a}. These are chemically rich regions in dense interstellar clouds which are warmed by an associated protostar and have typical densities of 10$^6$ -- 10$^8$ cm$^{-3}$ and temperatures on the order of 100 K. Observations suggest methyl formate abundances can be as high as $\sim$10$^{-8} \times n$(H$_2$) in these regions.

The standard view of the complex chemistry that pertains in hot cores requires that dust grains build up icy mantles at early times, when the temperature is low ($\sim$10 K) and the core is in a state of collapse \cite[]{brown88a}. Later, the formation of a nearby protostar quickly warms the gas and dust, re-injecting the grain mantle material into the gas phase, and stimulating a rich chemistry. Various hot core models \cite[]{millar91b,charnley92b,caselli93a,charnley95a} have shown that large oxygen-bearing molecules such as methyl formate may be formed during the ``hot'' stage when large amounts of methanol are released from the grain mantles. Methanol is easily protonated by H$_3^+$ ions, and the resulting protonated methanol was thought to react with formaldehyde to produce protonated methyl formate. Subsequent dissociative recombination with electrons would then complete the process. In this way, large amounts of methyl formate could be produced on timescales of 10$^4$ -- 10$^5$ yr after the evaporation of the ices.

However, this gas-phase mechanism has now become rather more contentious. Recent quantum chemical calculations carried out by \cite{horn04a} have shown that the first stage of the process -- reaction between H$_2$CO and CH$_3$OH$_2^+$ -- is highly inefficient at producing protonated methyl formate. This is due to a large potential barrier between isomeric complexes, resulting in an activation energy barrier for the reaction on the order of $\sim$15,000 K. Horn et al. suggested other gas-phase routes to produce protonated methyl formate as well as two other complexes, adopting moderate efficiencies for each to recombine with electrons to produce HCOOCH$_3$. Their chemical models (at 100 K) showed that even with optimistic rates, these reactions were unable to reproduce observed abundances. 

This considerable difficulty is compounded by increasing experimental evidence regarding the efficiencies of the dissociative recombinations of large saturated ions. \cite{geppert05a,geppert06a} have investigated the branching ratios of the recombination of protonated methanol, finding that the fraction of recombinations producing methanol is considerably lower than the 50 -- 100 \% 
typically assumed in chemical models \cite[see e.g.][]{leteuff00a}. \cite{geppert06a} find that just $3 \pm 2$ \% 
of recombinations result in methanol, with more than 80 \% 
resulting in separate C and O groups.  This work is in agreement with many previous storage-ring experiments on the products of the dissociative recombination reactions involving other H-rich species, which are dominated by three-body channels. The strong implication for such larger molecules as methyl formate and dimethyl ether, which were hitherto expected to form from their protonated ions, is that their production via recombination must be, at best, similarly inefficient. Therefore, the gas-phase routes to such molecules must be considered incapable of reproducing astronomically observed abundances at temperatures of $\sim$100 K.

Numerous allusions have been made \cite[e.g.][]{charnley95a,ehrenfreund00a,cazaux03a,peeters06a} to the possibility that methyl formate is produced on dust-grain surfaces at some point in the evolution of a hot core when dust temperatures are relatively low. The reaction set of \cite{allen77a}, a list of exothermic and plausible grain-surface reactions, includes one such association between the surface radicals HCO and CH$_3$O, producing HCOOCH$_3$. Many of the reactions listed in this set have been incorporated into the gas-grain chemical model of The Ohio State University (OSU) \cite[]{hasegawa92a}, but only very few of those involve more than one large radical. This is because at the low temperatures of dense interstellar clouds ($\sim$10 K), such large radicals would be effectively stationary on the grain surfaces, prohibiting their reaction with anything but small, relatively mobile species like atomic hydrogen. At the high temperatures encountered in hot molecular cores, these radicals would no longer be present on grain surfaces at all, having thermally evaporated. However, in the light of the failure of gas-phase chemistry to reproduce methyl formate observations, we must ask how these radicals should behave at intermediate temperatures. Clearly such a physical state must exist in the timeline of hot core evolution, but is it long enough to significantly influence the behaviour of the chemistry? Since most chemical models involve only a rudimentary treatment of grain surfaces, and implement only a gas-phase chemistry, few investigations of the warm-up phase of hot cores have been carried out. 

\begin{table}
\caption[]{Desorption energies of selected species}
\label{tab1}
\centering
\begin{tabular}{l c}
\hline\hline
\noalign{\smallskip}
Species & $E_D$ (K) \\
\noalign{\smallskip}
\hline
\noalign{\smallskip}
H & 450 \\
\noalign{\smallskip} 
H$_2$ & 430 \\
\noalign{\smallskip}
OH & 2850 \\
\noalign{\smallskip}
H$_2$O & 5700 \\
\noalign{\smallskip}
N$_2$ & 1000 \\
\noalign{\smallskip}
CO & 1150 \\
\noalign{\smallskip}
CH$_4$ & 1300 \\
\noalign{\smallskip}
HCO & 1600 \\
\noalign{\smallskip}
H$_2$CO  & 2050 \\
\noalign{\smallskip}
CH$_3$O  & 5080 \\
\noalign{\smallskip}
CH$_3$OH  & 5530 \\
\noalign{\smallskip}
HCOOH & 5570 \\
\noalign{\smallskip}
HCOOCH$_3$ & 6300 \\
\noalign{\smallskip}
CH$_3$OCH$_3$ & 3150 \\
\noalign{\smallskip}
\hline
\end{tabular}
\end{table}

\cite{viti99a} explored the effects of the selective, time-dependent re-injection of grain mantle-borne species, using the same thermal evaporation mechanism as employed in the OSU gas-grain code \cite[]{hasegawa92a}. They used a linear growth in  the temperature of the hot core, starting from 10 K, over periods corresponding to the switch-on times for hydrogen burning of variously sized protostars, as determined by \cite{bernasconi96a}. They investigated the effect of selective re-injection of mantle species on gas-phase chemical evolution. The follow-up work of \cite{viti04a} used results from temperature programmed desorption (TPD) laboratory experiments to determine temperatures at which particular species should desorb, taking into account phase changes in the icy mantles. Whilst these investigations shed new light on the dependence of the chemistry on (currently ill-defined) physical conditions, they necessarily ignored the effects of the grain-surface chemistry prior to re-injection.

In this study, we use the OSU gas-grain code to model the warm-up phase of hot-core evolution, with the primary aim of reproducing observed HCOOCH$_3$ abundances. The code was previously used to model hot cores by \cite{caselli93a}, requiring constant grain temperatures. The code is now capable of dealing with the variable temperatures we expect in the warm-up phase. The grain chemistry allows for time-dependent accretion onto and evaporation (thermal and cosmic ray-induced) from the grain surfaces. Surface-bound species may react together -- the majority of reactions are hydrogenations but reactions may take place between other atoms, and smaller radicals such as OH and CO, at rates determined by the grain temperature and diffusion barriers. Following the work of \cite{ruffle01a} we allow surface-bound molecules to be photodissociated by the (heavily extinguished) interstellar radiation field, and by the cosmic-ray induced field of \cite{prasad83a}. 

\begin{table*}
\caption[]{Reactions from \cite{horn04a}, with adjusted rates $^{a}$}
\label{tab2}
\centering
\begin{tabular}{l c c c}
\hline\hline
\noalign{\smallskip}
Reaction & $k(300$ K$)$        & $T$-dependence, $\alpha$ & $k(10$ K$)$          \\
\noalign{\smallskip}
         & (cm$^3$ s$^{-1}$) &                        & (cm$^3$ s$^{-1}$)  \\
\noalign{\smallskip}
\hline
\noalign{\smallskip}
(4) CH$_3$OH$_{2}^{+}$ + H$_2$CO $\rightarrow$ CH$_3$OH$_2$OCH$_{2}^{+}$ + $h \nu$ &  $3.1 \times 10^{-15}$  
&  -3    &  $8.4 \times 10^{-11}$ \\
\noalign{\smallskip} 
(5) H$_2$COH$^+$ + H$_2$CO $\rightarrow$ H$_2$COHOCH$_{2}^{+}$ + $h \nu$ &  $8.1 \times 10^{-15}$  
&  -3    &  $2.2 \times 10^{-10}$ \\
\noalign{\smallskip}
(6) CH$_{3}^{+}$ + HCOOH $\rightarrow$ HC(OH)OCH$_3^{+}$ + $h \nu$ &  $1.0 \times 10^{-11}$  
&  -1.5  &  $1.6 \times 10^{-9}$ \\
\noalign{\smallskip}
\hline
\noalign{\smallskip}
$^{a}$ $k(T) = k(300$ K$) \times (T/300)^{\alpha}$ \\
\end{tabular}
\\
\end{table*}

To this basic chemical model we add three new reactions from \cite{allen77a}, allowing for the formation of methyl formate, dimethyl ether (CH$_3$OCH$_3$) and formic acid (HCOOH) from grain-surface radicals. We employ a new set of desorption and diffusion barriers representing water ice. We also investigate the effects of the warm-up on the gas-phase reactions set out in \cite{horn04a}. Those reactions show a strong inverse temperature dependence, implying much larger rates at temperatures less than 100 K.

\section{Model}

\subsection{Physical Model}

We adopt a two-stage hot-core model: in stage 1, the gas collapses from a comparatively diffuse state of $n_H = 3 \times 10^{3}$ cm$^{-3}$, according to the isothermal free-fall collapse mechanism outlined in \cite{rawlings02a} and derived from \cite{spitzer78a}. This process takes just under 10$^6$ yr to reach the final density of $10^{7}$ cm$^{-3}$. 

In stage 2 the collapse is halted, and a gradual increase in dust and gas temperatures begins. We base this phase of our physical model on the approach of \cite{viti04a}. They used the observed protostellar luminosity function of \cite{molinari00a} to derive effective temperatures throughout the accretion phase of a central protostar, up until it reaches the zero-age main sequence (ZAMS), at which hydrogen burning takes place. They approximated the effective temperature profile as a power law with respect to the age of the protostar, then assumed that the temperature of the nearby hot core material should follow the same functional form. The contraction times determined by \cite{bernasconi96a} were used to provide a timescale for this process, on the order of 10$^4$ to 10$^6$ yr for stars of 60 to 5 solar masses, respectively. 

Here, we adopt power law temperature profiles for the hot core material, with a power index of either 1 (linear time-dependence) or 2. We allow the temperature to increase from 10 -- 200 K over three representative timescales: $5 \times 10^{4}$, $2 \times 10^{5}$ and $1 \times 10^{6}$ yr, approximating high-mass, intermediate, and low-mass star-formation, respectively. The high gas density of 10$^7$ cm$^{-3}$ implies a strong collisional coupling between the dust and gas temperatures, hence we assign them identical values at all times.

\subsection{Surface Chemistry}

For the grain-surface chemistry, we employ a new set of diffusion energy barriers and desorption energies for each species. Because the diffusion barriers determine the rate at which a species can hop between adjacent binding sites on the surface, the sum of these rates for two reaction partners determines the reaction rate \cite[]{hasegawa92a}. Similarly, desorption energies determine the rate of thermal and cosmic ray heating-induced desorption \cite[see][]{hasegawa93a}. Diffusion barriers are set as a fixed fraction of the desorption energy of each species, as in previous work \cite[e.g.][]{hasegawa92a,ruffle00a}. Both the diffusion rates and thermal desorption (evaporation) rates are exponentially dependent on the grain temperature, according to a Boltzmann factor.

Of the desorption energies that have been used in previous applications of the gas-grain code, some (including those for C, N, O, and S) correspond to a water ice surface, but most are given for carbonaceous or silicate surfaces. Here we construct a set of values corresponding to water ice for all the surface species in the gas-grain code, following \cite{cuppen06a}. These authors analysed various experimental data to obtain best values for H, H$_2$, O$_2$ and H$_2$O on non-porous ice. We use these values and add to them data provided by Dr. M. P. Collings (private communication), which correspond to the TPD results obtained by \cite{collings04a}. 
This gives us values for CO, CO$_2$, CH$_3$OH, C$_2$H$_2$, HCOOH, N$_2$, NH$_3$, CH$_3$CN, SO, SO$_2$, H$_2$S, and OCS. \cite[We ignore the value for CH$_4$, which was subject to experimental problems -- see][]{collings04a}. We additionally assume that the OH value is half that for H$_2$O, to take account of hydrogen bonding. We use these skeletal values (along with unchanged atomic values, except for H) and extrapolate to other species by addition or subtraction of energies from members of the set, making the assumption that the energies are the sum of the fragments that compose the molecules. The net result is that most species possess desorption barriers around 1.5 times larger than in our previous models, whilst those species that should exhibit hydrogen bonding to the surface can have barriers as much as five times as large. Table \ref{tab1} shows a selection of important surface species and their desorption energies, $E_D$. A complete table of energy parameters is available from the authors on request.

In previous models, diffusion barriers were set to either 0.3 or 0.77 of the desorption energies for most species (not atomic or molecular hydrogen); see \cite{ruffle00a}. Here we adopt a blanket value for all species of 0.5 $E_D$, based on an assortment of data on icy surfaces and our best estimate of the role of surface roughness and porosity.

\begin{table}
\caption[]{Initial abundances of H$_2$ and elements.}
\label{tab3}
\centering
\begin{tabular}{l c}
\hline\hline
\noalign{\smallskip}
Species $i$ & $n_{i}/n_{H}$ $^{\dag}$ \\
\noalign{\smallskip}
\hline
\noalign{\smallskip} 
H$_2$ & $0.5$ \\
\noalign{\smallskip}
He & $0.09$ \\
\noalign{\smallskip} 
C & $1.4(-4)$ \\
\noalign{\smallskip}
N & $7.5(-5)$ \\
\noalign{\smallskip}
O & $3.2(-4)$ \\
\noalign{\smallskip}
S & $1.5(-6)$ \\
\noalign{\smallskip}
Na & $2.0(-8)$ \\
\noalign{\smallskip}
Mg & $2.55(-6)$ \\
\noalign{\smallskip}
Si & $1.95(-6)$ \\
\noalign{\smallskip}
P & $2.3(-8)$ \\
\noalign{\smallskip}
Cl & $1.4(-8)$ \\
\noalign{\smallskip}
Fe & $7.4(-7)$ \\
\noalign{\smallskip}
\hline
\noalign{\smallskip}
$^{\dag}$ $a(b) = a \times 10^{b}$ \\
\end{tabular}
\end{table}

The reaction set we employ derives from that used by \cite{ruffle01a}, which includes grain-surface photodissociation. To that set we add the following three surface reactions from \cite{allen77a}:
\smallskip
\begin{equation}
\mbox{HCO} + \mbox{CH}_{3}\mbox{O} \rightarrow \mbox{HCOOCH}_{3},
\end{equation}
\begin{equation}
\mbox{CH}_{3} + \mbox{CH}_{3}\mbox{O} \rightarrow \mbox{CH}_{3}\mbox{OCH}_{3},
\end{equation}
\begin{equation}
\mbox{HCO} + \mbox{OH} \rightarrow \mbox{HCOOH}.
\end{equation}
\smallskip

These reactions we deem to be the most likely surface routes to each of the three molecules. During the cold phase, the radicals HCO and CH$_3$O are formed by successive hydrogenation of CO, which mostly comes directly from accretion from the gas phase. The reactions H + CO $\rightarrow$ HCO and H + H$_2$CO $\rightarrow$ CH$_3$O each assume an activation energy of 2500 K \cite[]{ruffle01a}. Large reserves of H$_2$CO and CH$_3$OH may build up through the hydrogenation process. However, at warmer temperatures such processes should be less efficient, due to the faster thermal evaporation of atomic hydrogen. Under these conditions, processes which break down H$_2$CO and CH$_3$OH may become important.

\subsection{Modified Rates}

Applications of the gas-grain code to cold dark clouds in previous publications have made use of ``modified'' rates in certain situations; specifically, problems arose when solving for species with average grain-surface abundances on the order of unity or below \cite[]{caselli98a}. Comparison with results from simple Monte Carlo techniques \cite[]{tielens82a} indicated that atomic hydrogen in particular reacted too quickly when using the rate-equation method employed in the gas-grain code. In effect, the hydrogen was reacting more quickly than it was actually landing on the grain surfaces. This problem was fixed by limiting the rate of diffusion of the hydrogen to no more than the rate of accretion. Additionally, when in the ``evaporation limit'' (i.e. when the rate of evaporation exceeds the rate of accretion), the diffusion rate was limited to the rate of H evaporation instead, since in this case it is the evaporation rate which is the main determinant of surface hydrogen abundance. Whilst \cite{shalabiea98a} applied the modified rate technique to all species, the effect was found to be small except for atomic H, at the low temperatures of dark clouds ($\sim$$10$ K). The analysis by \cite{katz99a} of earlier experimental work subsequently showed that surface hydrogen hops between binding sites thermally, rather than via quantum tunnelling, indicating a much reduced diffusion rate for atomic hydrogen. The use of this result by \cite{ruffle00a,ruffle01a} showed that the need for modified rates even for atomic hydrogen was marginal.

\begin{figure*}
\centering
\includegraphics[width=8.5cm]{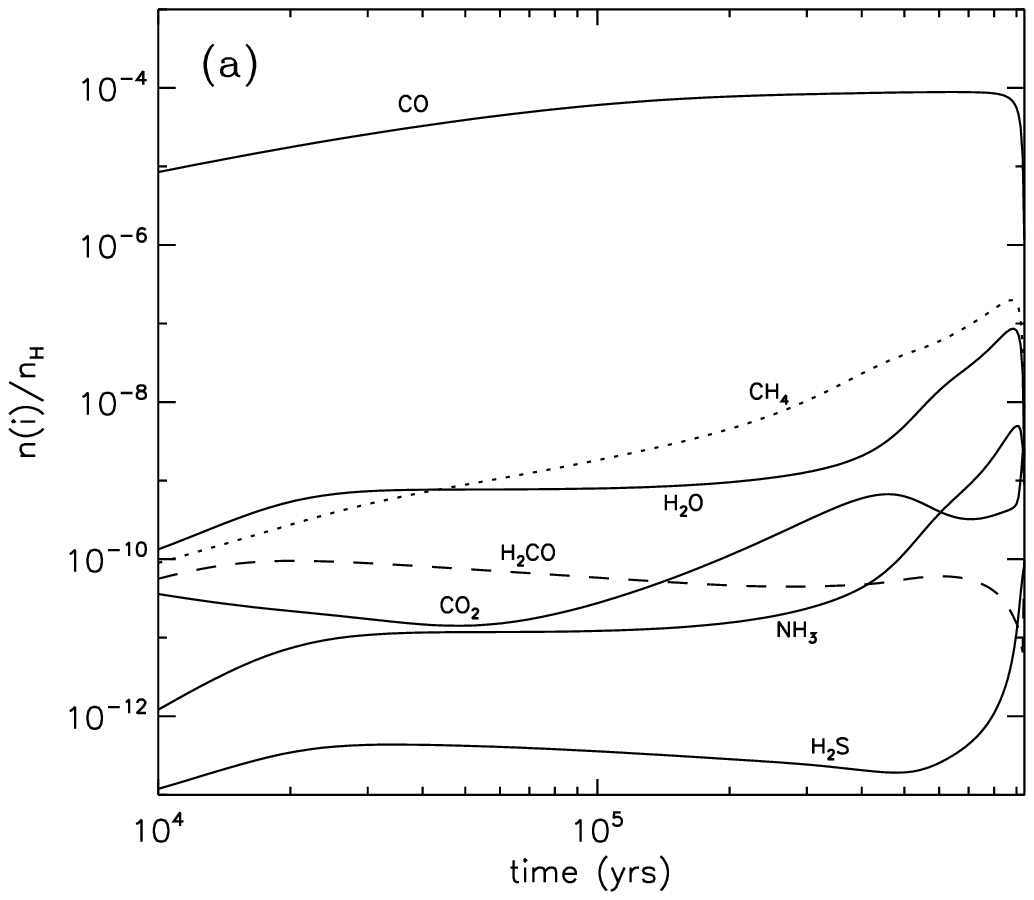}
\includegraphics[width=8.5cm]{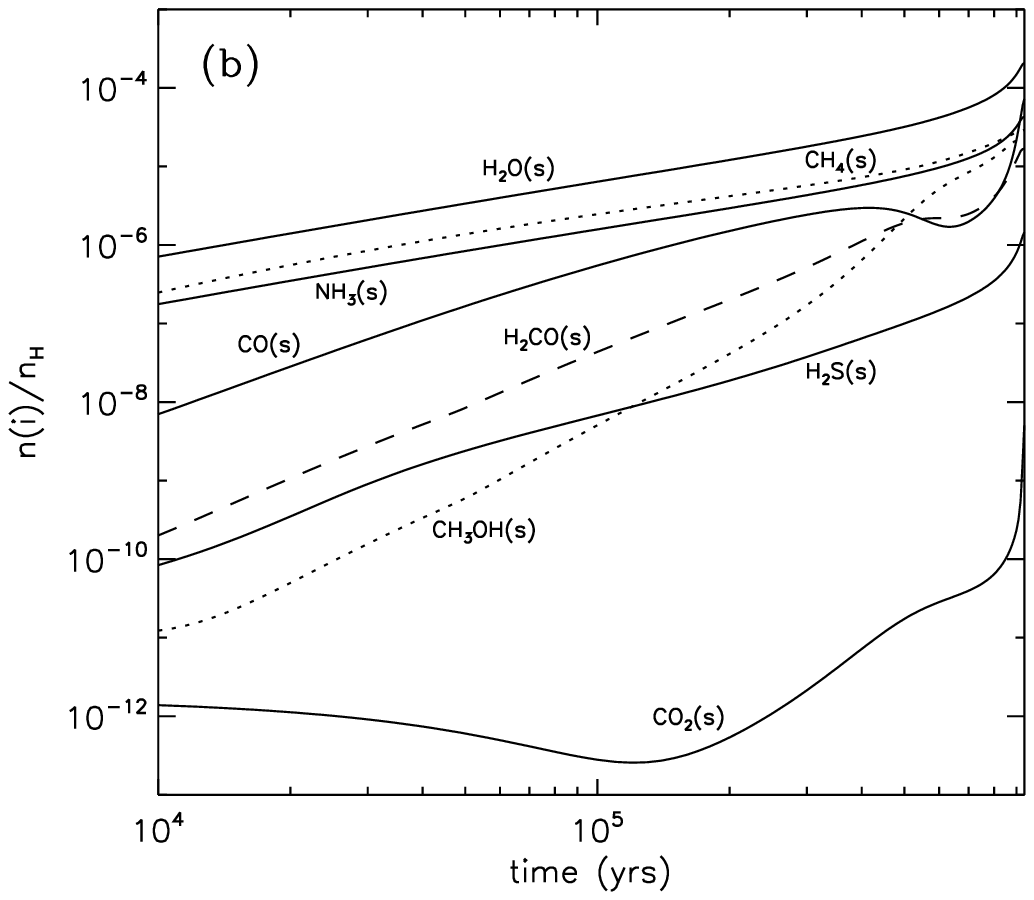}
\caption{Chemistry during the collapse phase. Surface species are designated by (s).}
\label{figure1}
\end{figure*}

However, in this work, we deal with much higher temperatures -- at which even the heaviest species can become highly mobile. This means that any species could react at a faster rate than it is accreted or evaporated when some critical temperature is reached. But this does not necessarily mean that modified rates should be invoked for all species and reactions. Many of the heavy radicals in which we are interested here are formed mainly by surface reactions, and are therefore not governed by accretion in the way that many species are whilst at low temperatures. Also, some species can be produced by cosmic ray photodissociation of an abundant surface molecule. These complications suggest that any need to apply the modified rates is highly specific, both to individual species and to particular brief periods for those species. Added to this is the fact that the modified rates are untested against Monte Carlo codes at temperatures much greater than 20 K, meaning that there is no certain need to modify the rates at all in those cases. 

Hence, we modify only reactions involving atomic hydrogen. It is the only species that requires modification at less than 20 K, and is easily evaporated above 20 K, making it less important in the surface chemistry at such temperatures. As a result,  the modified rates mainly come into play in the low-temperature collapse phase. The rates of surface reactions involving atomic hydrogen are modified in the manner outlined by \cite{ruffle00a}. Additionally, we modify the rates of hydrogen reactions that have activation energy barriers, following the method of \cite{aikawa05a}. This addition is necessary because at the high densities involved, the quantities of reactant that build up can become large enough that, even with the presence of the activation energy barrier, the rate at which hydrogen reacts is greater than the rate at which it is being deposited on the surfaces. To avoid such a run-away hydrogen chemistry, if the product of the activation barrier term, the surface abundance of the (non-hydrogenic) reactant, and the diffusion rate, is greater than the faster of the rates of accretion and evaporation of atomic hydrogen, then that product is replaced in the rate equation with the accretion or evaporation rate. For a full discussion of this issue see \cite{caselli02a}.

\subsection{Gas-Phase Chemistry}

Table \ref{tab2} shows reactions previously included in \cite{horn04a} that lead, without activation energy, to protonated methyl formate and to two other methyl formate precursors. Two of the ionic products require significant structural re-arrangement in a subsequent dissociative recombination reaction to produce methyl formate.  The rate coefficient given to the first reaction in the table (labelled 4) is unchanged; to the other two (5 and 6) we attribute somewhat weaker temperature dependences due either to competition or a better estimate. At a temperature of 10 K, all three reactions are significantly faster than at the 100 K at which they were tested by Horn et al.; reaction (6) reaches approximately the collisional (i.e. maximum) rate coefficient at this temperature. Hence, we expect contributions from these reactions to be strong during the early stages of the warm-up phase, if the abundances of the reactants are significant. We allow the products of reaction (6) to recombine with electrons to form methyl formate in 5 \% 
of reactions, two times lower than assumed by Horn et al. The protonated products of reactions (4) and (5) are allowed to form methyl formate in just 1 \% 
of recombinations, because each of these channels would require a more drastic structural change within the complex.

We also include approximately 200 gas-phase reactions from the UMIST ratefile \cite[]{leteuff00a} which were not present in the (predominantly low temperature) OSU reaction set. These reactions have activation energies, but the greater temperatures seen in hot cores justify their inclusion here.

The initial gas-phase matter is assumed to consist of atoms with the exception of molecular hydrogen. Shown in table \ref{tab3}, these values derive from those selected by \cite{wakelam06a} from the most recent observational elemental determinations for diffuse clouds. Since the high temperatures attained in hot cores are capable of driving off the entirety of the granular icy mantles, we choose relatively undepleted (i.e. diffuse cloud) values for the lighter elements, in contrast with previous applications of the gas-grain code \cite[most pertinently,][]{caselli93a}. We allow heavier elements to be depleted by an order of magnitude, on the assumption that the remainder is bound within the grain nuclei. This also ensures that fractional ionizations stay at appropriate levels, defined by \cite{mckee89a} as $X(e) \simeq 10^{-5} n($H$_{2})^{-1/2}$. The behaviour of sulphur on grain surfaces is not fully understood, and the form it takes on/in the grains is not known; observations have so far failed to detect H$_2$S on grains, and some authors suggest OCS may be the dominant form \cite[]{vandertak03a}. Moreover, the levels of sulphur-bearing species detected in the gas phase of hot cores are typically not as high as the levels detected in diffuse clouds \cite[see e.g.][]{wakelam04a}. We therefore deplete sulphur by an order of magnitude from the values of \cite{wakelam06a}.

\section{Results}

\subsection{Stage 1: Collapse}

All models investigated here use the same stage 1 collapse model, hence the initial conditions for the warm-up phase are the same in all cases. Figure \ref{figure1} shows the evolution of some important gas-phase and grain-surface species, with respect to $n_H$, during the collapse from $3 \times 10^{3}$ -- $10^{7}$ cm$^{-3}$.

\begin{figure}
\centering
\includegraphics[width=8.5cm]{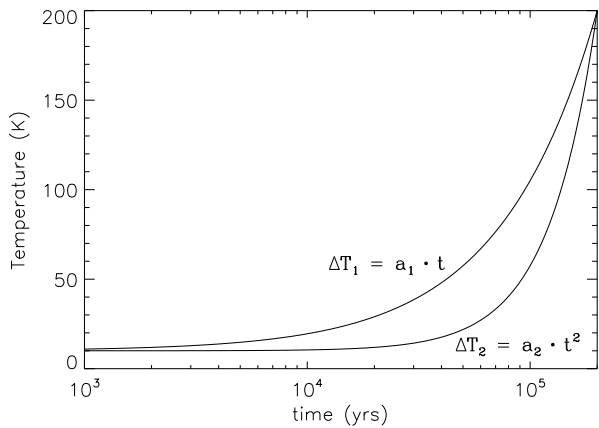}
\caption{Temperature profiles for a protostellar switch-on time of $t=2 \times 10^{5}$ yr, adopting linear ($T_1$) and $t^2$ time dependences ($T_2$).}
\label{figure2}
\end{figure}

Initially, the collapse is slow, and most of the significant increase in density takes place during the final $5 \times 10^{5}$ yr. Over this period, we see that gas-phase reactions become much faster due to the greater density, whilst accretion rates also increase, allowing for an accelerated surface chemistry. Large amounts of H$_2$O, CO, CH$_4$, H$_2$CO, CH$_3$OH, H$_2$S and NH$_3$ have built up on the grain surfaces by the end of the collapse. Besides water ice, NH$_3$ and CH$_4$ make up a large and fairly constant proportion of the ice. CH$_4$ ranges from $\sim$10 -- 24 \%
of total ice throughout the collapse phase; NH$_3$ is steady at around 15 \%. 
Water, methane and ammonia ices are formed by successive hydrogenation of O, C and N respectively. Carbon monoxide is predominantly formed in the gas phase and then accreted onto surfaces. At the very end of the collapse, CO deposition outpaces H$_2$O formation, due to the shortened supply of atomic oxygen. The model suggests that the outer 50 or so monolayers of ice should have a CO:H$_2$O ratio of approximately 6:4. Formaldehyde, which is formed by hydrogenation of CO, should comprise around 5 -- 10 \%
of these outer layers.

The calculated abundances of ices at the end of the collapse are in reasonable agreement with IR observations \cite[e.g.][]{nummelin01a} except for CO$_2$, which is greatly underproduced at 10 K. Solutions to this problem have been advanced by \cite{ruffle01b}. Observed CO$_2$ percentages (in dark cloud ices) are typically the same as or a little less than CO levels. Hence, even with efficient CO$_2$ production, there should not be a drastic effect on CO, H$_2$CO, or CH$_3$OH ice abundances in the cold phase. Once formed, CO$_2$ does not take part in further reactions, but may be photodissociated.

\subsection{Stage 2: Warm-up Phase}

\begin{figure*}
\centering
\includegraphics[width=8.5cm]{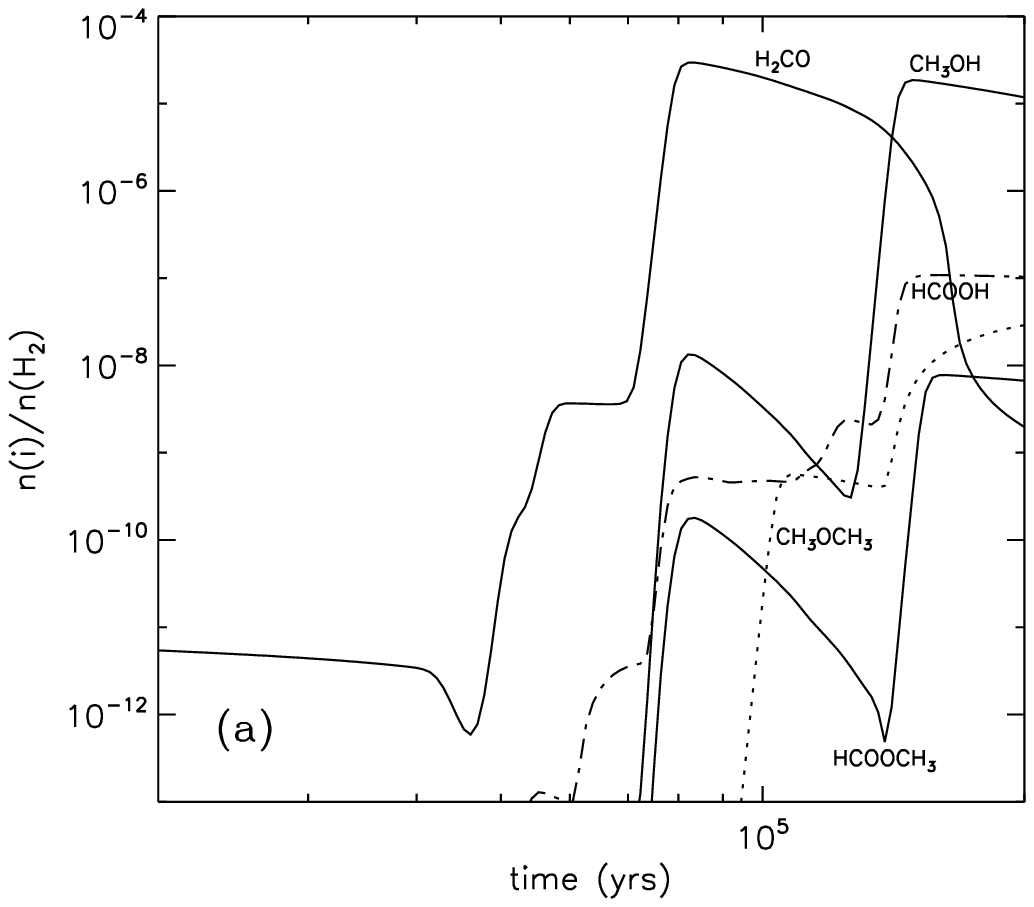}
\includegraphics[width=8.5cm]{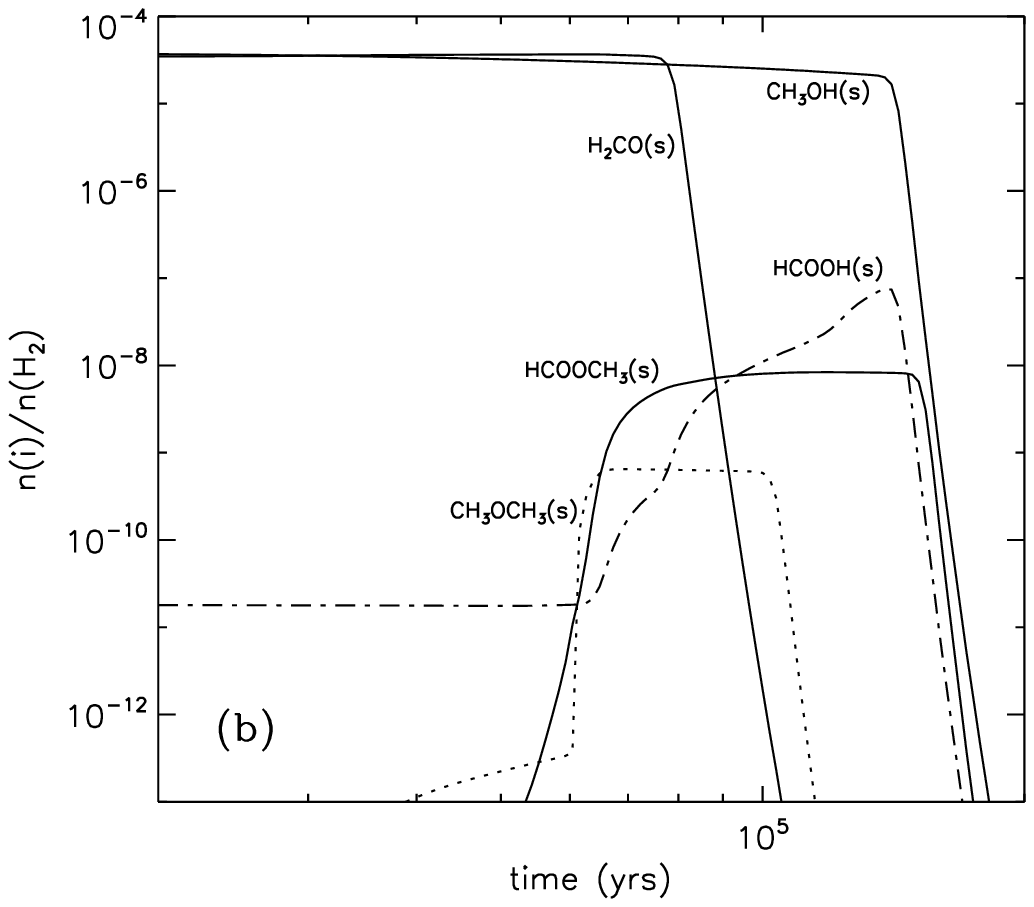}
\includegraphics[width=8.5cm]{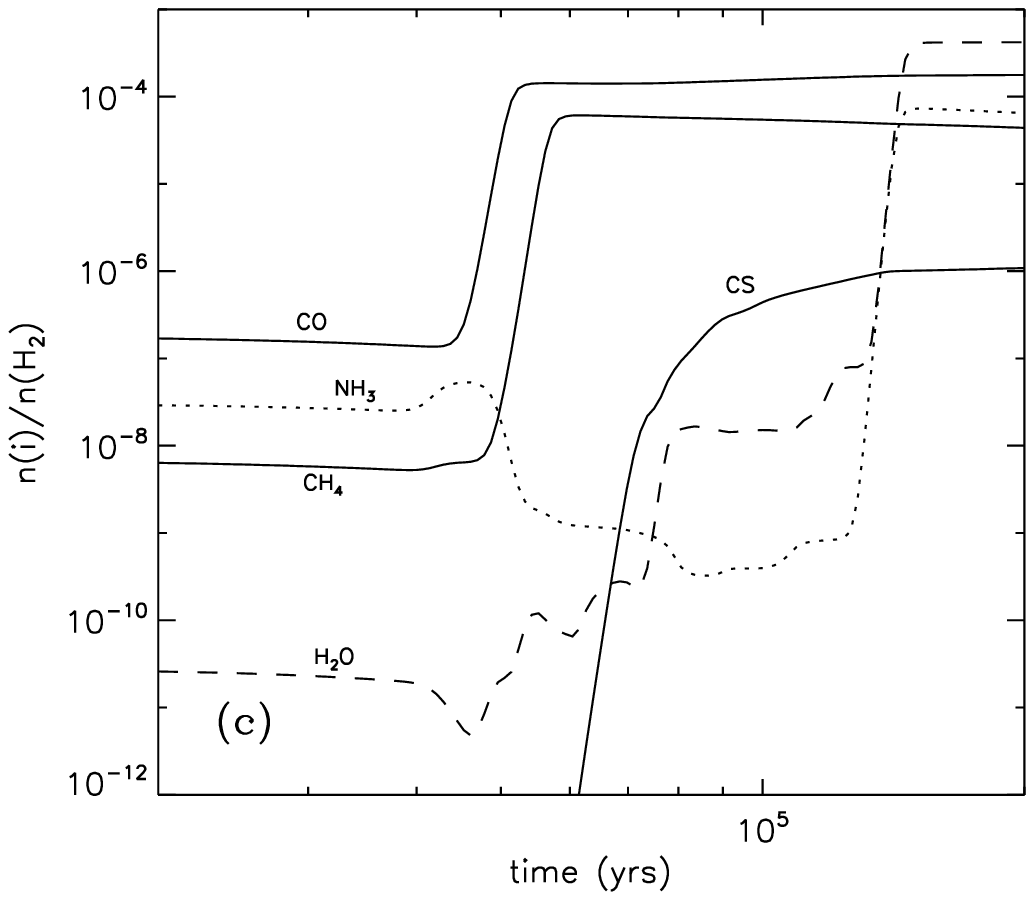}
\includegraphics[width=8.5cm]{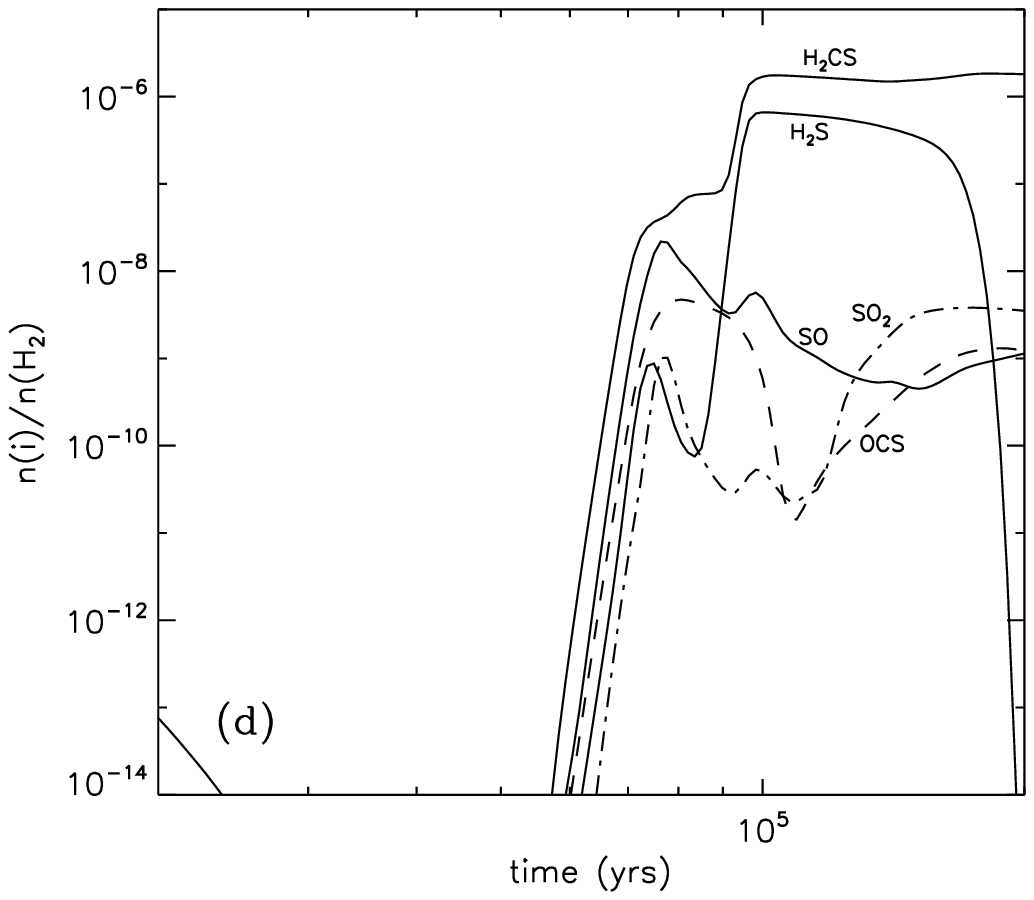}
\caption{Chemistry during standard warm-up phase -- $t_{f}=2 \times 10^{5}$ yr, $T=T_{2}(t)$. Surface species are marked (s). All abundances are shown with reference to molecular hydrogen.}
\label{figure3}
\end{figure*}

We first explore the behaviour of a ``standard'' physical and chemical model, before discussing other variations and free parameters. Figure \ref{figure2} shows two temperature profiles used in the warm-up phase of the model; our ``standard'' model uses the $\Delta T_2$ profile with a protostellar switch-on time of $2 \times 10^{5}$ yr, corresponding to the final time acheived in the model, $t_f$.  All of reactions (1) -- (6) are enabled. Figure \ref{figure3} shows plots of pertinent gas-phase and grain-surface species over the final ``decade'' of evolution; i.e. $2 \times 10^{4}$ -- $2 \times 10^{5}$ yr. Over this period, temperatures begin to rise significantly above 10 K, and ultimately reach the final value of 200 K.

\subsubsection{Methyl Formate, Formic Acid, and Dimethyl Ether}

Methyl formate is formed in significant quantities, both in the gas phase and on grain surfaces. Figure \ref{figure3}a shows the gas-phase abundance at late times -- a double-peaked structure is apparent. The first, lesser, peak comes about solely from gas-phase formation via reaction (5). Formaldehyde injected into the gas phase from the grain surfaces at this time is protonated by reaction with H$_3$$^+$ and HCO$^+$ ions. At $t=8 \times 10^{4}$ yr, a temperature of $\sim$40 K obtains, giving a rate coefficient for reaction (5) of $3.4 \times 10^{-12}$ cm$^3$ s$^{-1}$. This is more than an order of magnitude larger than the rate at 100 K \cite[cf.][]{horn04a}, a temperature typical of hot core chemical models. These factors combine to produce just under 10$^{-8}$ with respect to H$_2$ of methyl formate; however, this abundance of HCOOCH$_3$ does not manifest itself fully in the gas phase, since most of it is quickly accreted onto the grains. (The high density produces an accretion timescale of $\sim$2000 yr). This causes the sharp fall-off in gas-phase abundance. Reactions (4) and (6) (see table \ref{tab2}) do not contribute significantly to methyl formate production at any stage, firstly because their reactants are not typically present in large quantities at the same times; and secondly, because when the reactants H$_2$CO, HCOOH, and the methanol required as a precursor for reaction (4) are all abundant, temperatures are much higher (around 100 K), and the association rates are consequently much lower.  The second, final, peak in the gas-phase methyl formate abundance arises from the evaporation of the surface species.  

Figure \ref{figure3}b plots the grain-surface abundance of HCOOCH$_3$. Although not perceptible from the plot, there is a subtle shift in the formation route as the methyl formate abundance grows. The initial rise at around  $5 - 6 \times 10^{4}$ yr is brought about by the surface reaction (1) between the radicals HCO and CH$_3$O. This rise takes place when CO has strongly evaporated, whilst the elevation of the grain-surface temperature increases the mobility of OH and H$_2$CO, causing the rate of the reaction

\smallskip
\noindent OH + H$_{2}$CO $\rightarrow$ HCO + H$_{2}$O
\smallskip

\noindent to greatly increase. (The OH radical is produced by the photodissociation of water ice, initiated by cosmic ray-induced photons). The mobility of the resultant HCO has also greatly increased. At these warmer temperatures, the hydrogenation of CO has ceased to be important to HCO formation, due to the lower abundance of atomic H (caused by its faster thermal evaporation). However, when CO strongly evaporates at $\sim 20$ K, the reaction CO + OH $\rightarrow$ CO$_2$ + H -- the major destruction route of OH -- is removed. This greatly increases the OH abundance, allowing much HCO to be formed. At this point in time, although atomic H is more scarce than at lower temperatures, the reaction

\smallskip
\noindent H + H$_{2}$CO $\rightarrow$ CH$_{3}$O,
\smallskip

\noindent is still the dominant mechanism for CH$_{3}$O formation. H$_2$CO is abundant on the grains, having built up during the collapse phase. When significant amounts of H$_2$CO begin to evaporate, at $\sim$$8 \times 10^{4}$ yr ($T \simeq 40$ K), HCO and CH$_{3}$O production is reduced and surface formation of HCOOCH$_3$ is inhibited. At this point, reaction (5) starts to have a strong effect due to the H$_2$CO now present in the gas phase, and much of the methyl formate produced is then accreted. From this time until HCOOCH$_3$ evaporates, the two processes operate in tandem, with the gas-phase/accretion mechanism dominant but gradually decreasing in influence. The gas-phase/accretion route is responsible for the formation of around 25 \% 
of the total HCOOCH$_3$ present on the grain surfaces before evaporation. 

\begin{figure*}
\centering
\includegraphics[width=8.5cm]{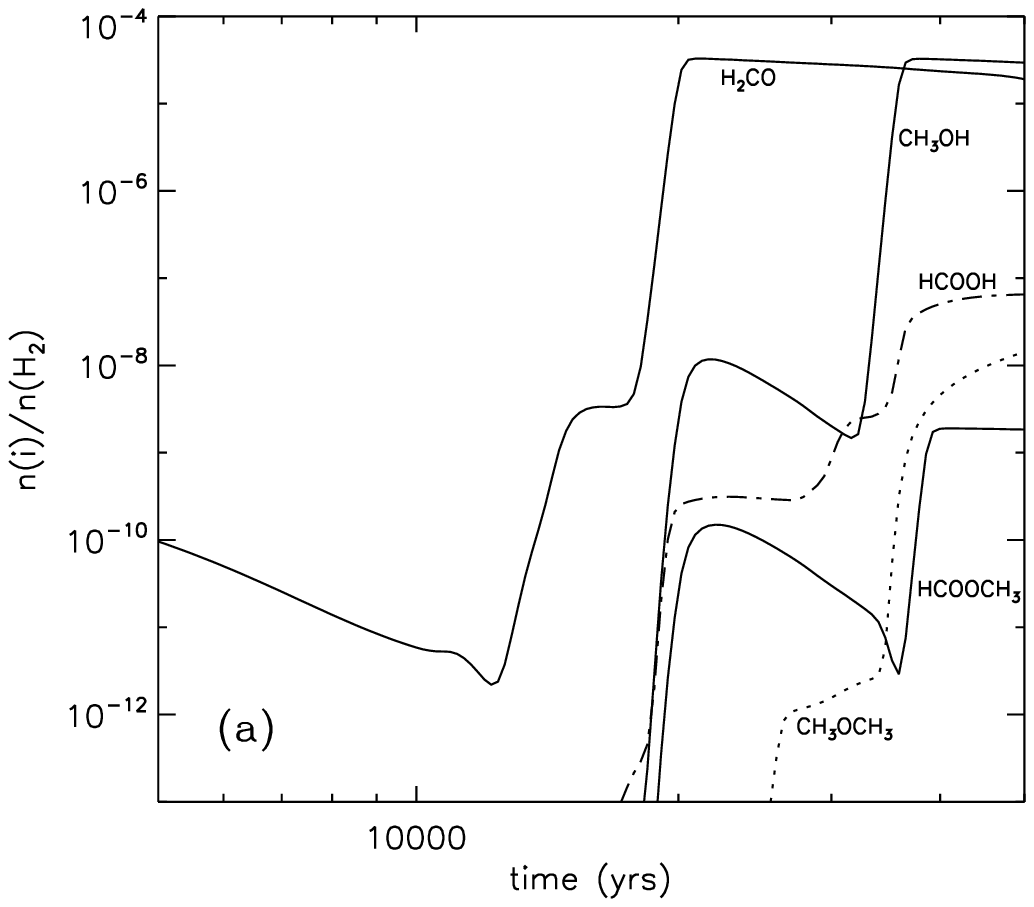}
\includegraphics[width=8.5cm]{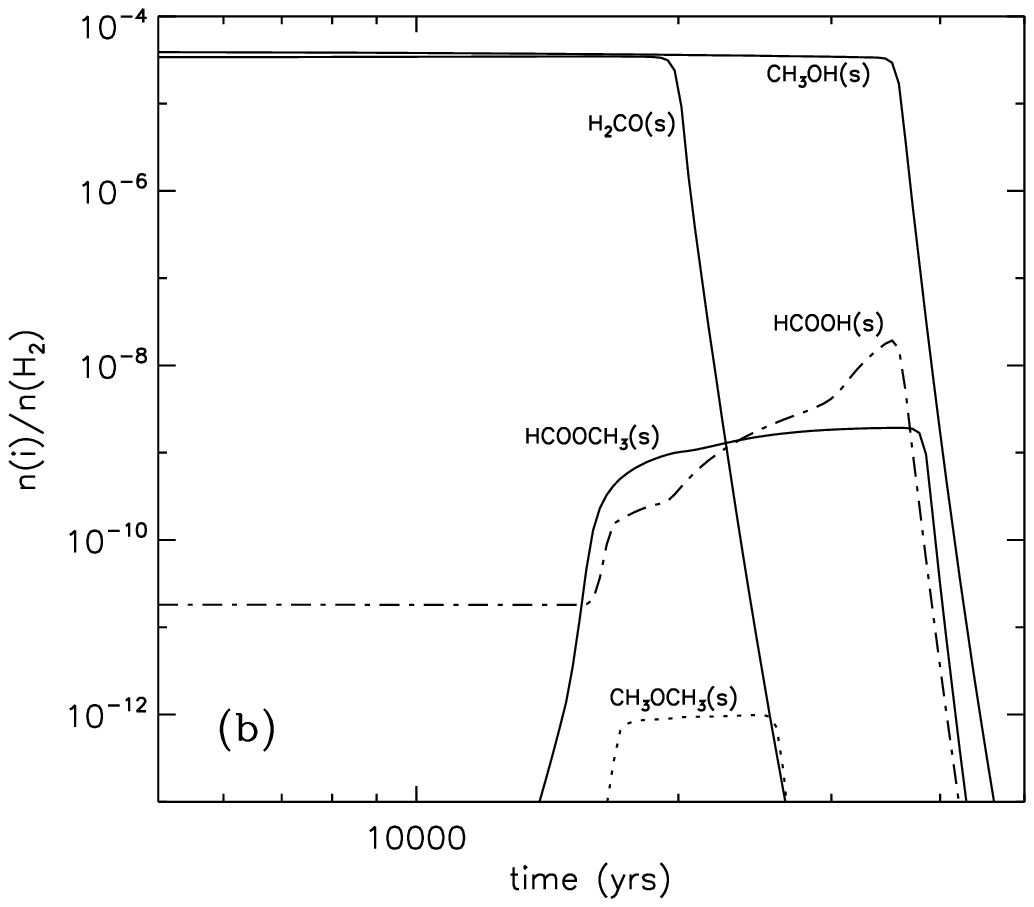}
\caption{Chemistry during short warm-up phase -- $t_{f}=5 \times 10^{4}$ yr, $T=T_{2}(t)$. Species marked (s) are surface species. All abundances are shown in reference to molecular hydrogen.}
\label{figure4}
\end{figure*}

At early times, HCOOH is formed in the gas phase from the dissociative recombination of HCOOH$_2^+$, which forms by radiative association of HCO$^+$ and H$_2$O. However, when CH$_4$ evaporates at around $5 - 6 \times 10^{4}$ yr (see figure \ref{figure3}c for the gas-phase profile), it can react with O$_2^+$ ions, forming HCOOH$_2^+$, which  then dissociatively recombines with electrons. This process results in the rise in the gas-phase formic acid abundance beginning at $6 \times 10^{4}$ yr. At the same time, as with methyl formate, HCOOH surface formation benefits from the increased formation, and diffusion rate, of HCO, causing the first rise in surface abundance.

The next rise in the gas-phase HCOOH abundance comes about when formaldehyde evaporates, which facilitates more OH formation in the gas phase via reaction with atomic oxygen. The OH radical and H$_2$CO then react in the gas phase to form HCOOH via the process

\smallskip
\noindent OH + H$_{2}$CO $\rightarrow$ HCOOH + H.
\smallskip

\noindent  Both of these gas-phase mechanisms increase the grain-surface HCOOH abundance by subsequent accretion of the product. 

On the grain surfaces, reaction (3) becomes most important when formaldehyde has evaporated significantly, because previously most OH was destroyed by reaction with H$_2$CO. After its evaporation, formaldehyde may still accrete and then (if it does not immediately evaporate again) it may react with OH to form HCO. Sufficient OH is still available to further react with HCO via reaction (3) because the surface abundance of H$_2$CO is comparatively small. At this stage, the dominant OH formation mechanism is photodissociation of H$_2$O by cosmic ray photons. The post-evaporation surface abundance of formaldehyde is small compared to its pre-evaporation level, but the gas-phase abundance remains high, and so accretion is fast and remains steady. Hence, we find that the formation of formic acid on surfaces is optimised at a temperature where the formaldehyde does not stick to the grains ($\sim$40 K), but resides in the gas phase and accretes and re-evaporates quickly. This allows it to react with OH to form enough HCO for reaction (3), whilst not dominating the OH chemistry to such an extent that reaction (3) suffers. Reaction (3) requires HCO to be formed at a significant rate, since HCO quickly evaporates at such temperatures. This optimum state for reaction (3) ends when the temperature increases to the point where OH evaporates ($\sim$70 K). 

However, the grain-surface formation mechanism is never dominant; the gas-phase synthesis (OH + H$_2$CO) is still around ten times faster at this point, due to high formaldehyde abundances. The final peak in surface HCOOH abundance and the small blip before the final gas-phase peak are due to increased gas-phase formation as OH evaporates from the grains.
\begin{figure*}
\centering
\includegraphics[width=8.5cm]{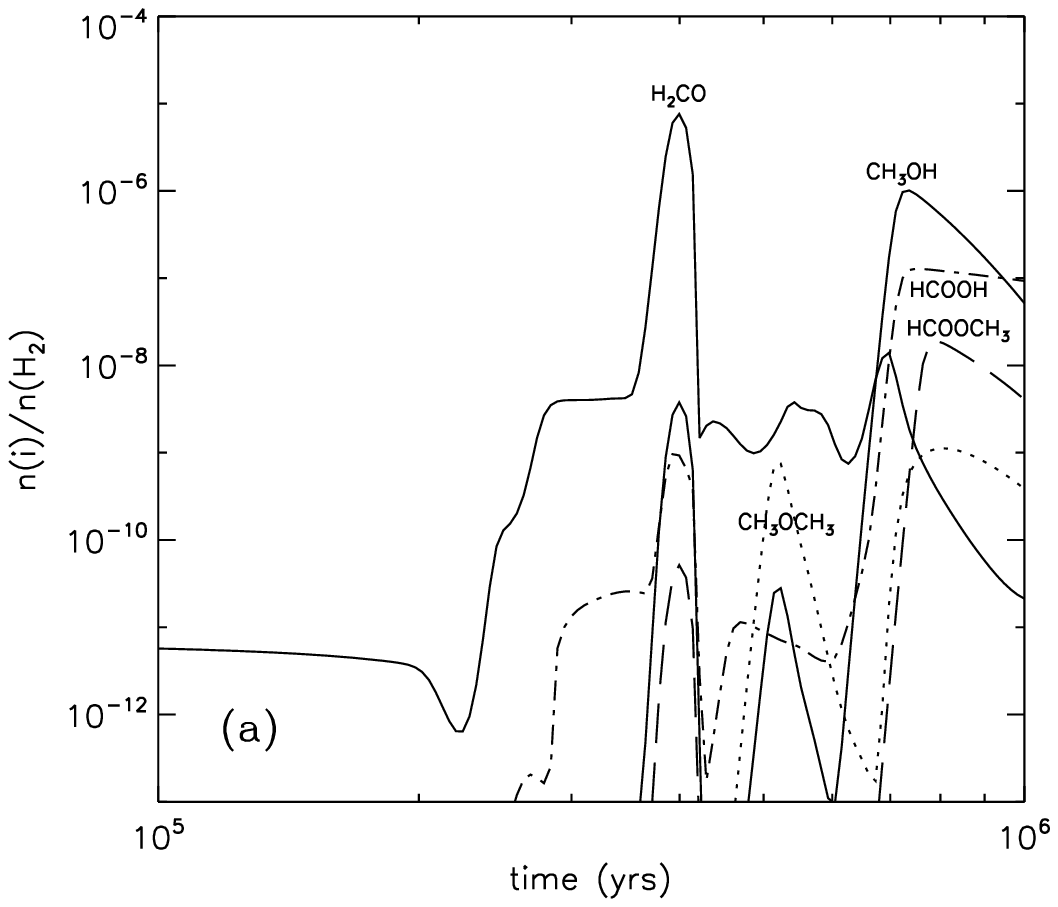}
\includegraphics[width=8.5cm]{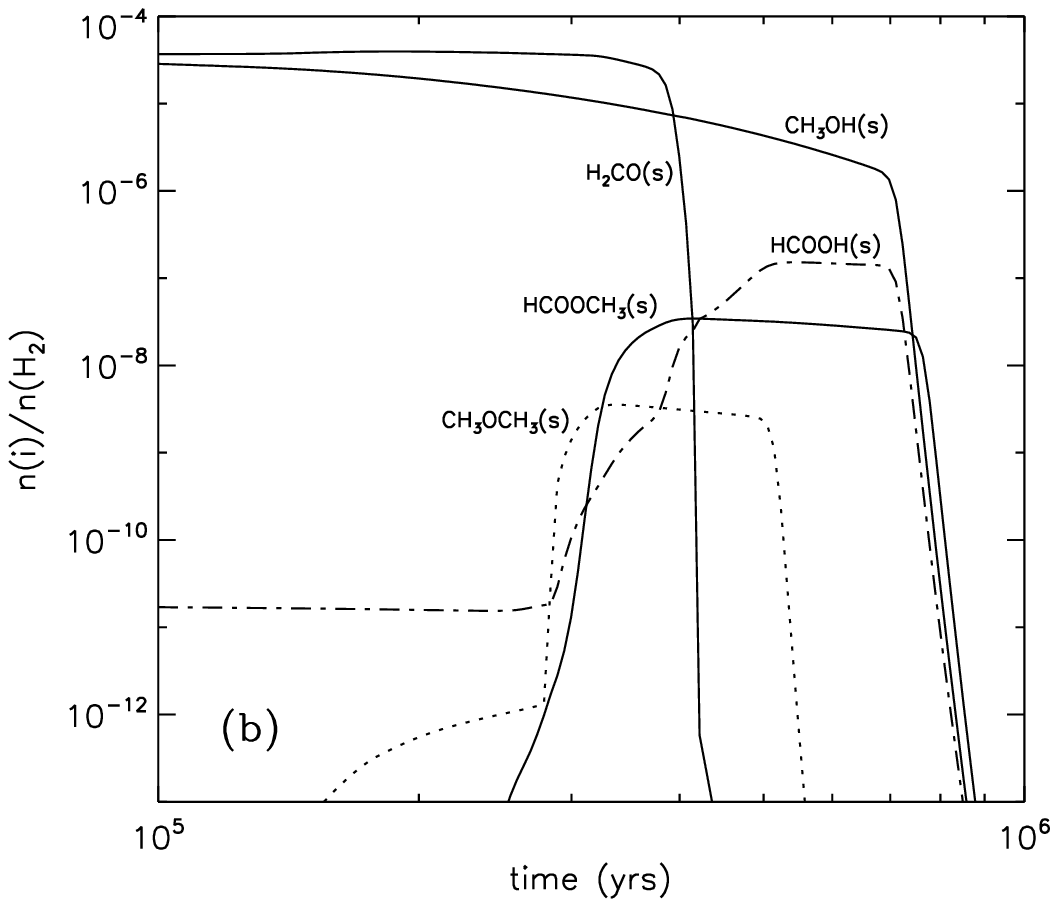}
\caption{Chemistry during long warm-up phase -- $t_{f}=1 \times 10^{6}$ yr, $T=T_{2}(t)$. Species marked (s) are surface species. All abundances are shown with reference to molecular hydrogen.}
\label{figure5}
\end{figure*}
Finally, the surface-bound HCOOH evaporates, leaving a high gas-phase abundance of $\sim 10^{-7}$ with respect to H$_2$. The majority of HCOOH is formed by the gas-phase reaction.

Dimethyl ether (CH$_3$OCH$_3$) is formed initially on grain surfaces by reaction (2). The strong rise up to peak surface abundance begins when the surface HNO abundance falls. This species is until then the main reaction partner of CH$_3$, forming CH$_4$ and NO. CH$_3$ is formed on the surfaces by the hydrogenation of CH$_2$ following the photodissociation of CH$_4$ by cosmic ray-induced photons. The fall in HNO allows CH$_3$ abundances to rise, fuelling reaction (2). Dimethyl ether continues to be formed by the grain-surface mechanism until it evaporates from the grains.

Reaction (2) does not contribute much dimethyl ether abundance compared to the final level, and this molecule evaporates earlier than methyl formate or formic acid, at a time of $\sim 10^{5}$ yr. The main formation mechanism is therefore the gas-phase route, which consists of the reaction between methanol and protonated methanol, which is fast when methanol finally evaporates, followed by dissociative recombination. We use a recombination efficiency for producing dimethyl ether of $\sim$1.5 \%.

\subsubsection{Other Species}

The gas-phase profile for H$_2$CO (figure \ref{figure3}a) also shows a complicated structure. Again, this is due to the time dependence of the evaporation of different species. The first dip (at $\sim 4.5 \times 10^{4}$ yr) is due to N$_2$ evaporation, which contributes strongly to the NH$_3$ gas-phase abundance; the ammonia competes with atomic oxygen to react with CH$_3$, which is required to form H$_2$CO in the gas phase at that time. The subsequent evaporations of CO and CH$_4$ then push up formaldehyde abundance strongly. The final rise in gas-phase abundance takes place as formaldehyde itself is desorbed from the grain surfaces.

The profiles of sulphur-bearing species resulting from this treatment show some features of interest. Importantly, H$_2$CS is the most abundant sulphur-bearing species at late times, in spite of the dominance of H$_2$S on the grain surfaces before this. When H$_2$S evaporates, it is  broken down to atomic sulphur by cosmic ray-induced photons. Meanwhile, CH$_3$ is abundant, both as a result of the high CH$_4$ level in the gas phase, and the cosmic ray-induced photodissociation of grain-surface methanol followed by the fast evaporation of the resultant CH$_3$. Atomic S and CH$_3$ then react in the gas phase to produce H$_2$CS. When methanol evaporates, CH$_3$ may still be formed by cosmic ray-induced photodissociation, but also by dissociative recombination of CH$_3$OHCH$_3^+$. The final H$_2$CS abundance is rather larger than observations might suggest, and larger than other models predict. Van der Tak et al. (2003) \nocite{vandertak03a} suggest values of 10$^{-9}$, using radiative transfer models with observations of high mass star-forming regions. Typical H$_2$CS:H$_2$S values are on the order of a few. However, that study suggests SO and SO$_2$ should be of the same order as H$_2$S and H$_2$CS, which is not the case here, except at times of around $8 \times 10^{4}$ yr. The observational study of \cite{wakelam04b} provides SO, SO$_2$ and H$_2$S values on the order of 10$^{-6}$. 

However, the sulphur chemistry is complex, and is not the main focus of this paper. We direct the reader to \nocite{viti06a} Viti, Garrod \& Herbst (2006, in prep.), which builds on the sulphur chemistry analysis of \cite{viti04a}, and includes grain-surface chemistry.

\begin{table*}
\caption[]{Selected gas-phase fractional abundances with respect to H$_2$, determined at the time of maximum HCOOCH$_3$ abundance.$^{a}$}
\label{tab4}
\centering
\begin{tabular}{l c c c c}
\hline\hline
\noalign{\smallskip}
Species   &  \multicolumn{4}{c}{Model} \\
\noalign{\smallskip}
\cline{2-5}
\noalign{\smallskip}
              & $t_{f}=5 \times 10^{4}$ yr & $t_{f}=2 \times 10^{5}$ yr & $t_{f}=1 \times 10^{6}$ yr & $t_{f}=2 \times 10^{5}$ yr \\
\noalign{\smallskip}
              & $T=T_2(t)$ & $T=T_2(t)$ & $T=T_2(t)$ & $T=T_1(t)$ \\
\noalign{\smallskip}
\hline
\noalign{\smallskip}
HCOOCH$_3$ (all reactions)       & 1.9$(-9)$  & 7.8$(-9)$  & 1.9$(-8)$  & 5.3$(-9)$ \\
\noalign{\smallskip}
HCOOCH$_3$ (no gas-phase routes) & 1.0$(-10)$ & 5.5$(-9)$  & 1.9$(-8)$  & 3.2$(-9)$ \\
\noalign{\smallskip}
HCOOH                            & 5.5$(-8)$  & 1.1$(-7)$  & 1.2$(-7)$  & 1.2$(-7)$ \\
\noalign{\smallskip} 
CH$_3$OCH$_3$                    & 6.1$(-9)$  & 1.4$(-8)$  & 1.1$(-9)$  & 2.2$(-8)$ \\
\noalign{\smallskip} 
H$_2$CO                          & 2.3$(-5)$  & 2.3$(-7)$  & 3.9$(-10)$ & 1.5$(-8)$ \\
\noalign{\smallskip} 
CH$_3$OH                         & 3.2$(-5)$  & 1.7$(-5)$  & 5.9$(-7)$  & 1.8$(-5)$ \\
\noalign{\smallskip}
\hline
\noalign{\smallskip}
$^{a}$ $a(b) = a \times 10^{b}$ \\
\end{tabular}
\end{table*}

\subsection{Timescale and Temperature Profile Variations}

In addition to the standard switch-on time of $2 \times 10^{5}$ yr, we ran additional models with heat-up times of $5 \times 10^{4}$ and $1 \times 10^{6}$ yr.  Figures \ref{figure4} and \ref{figure5} show gas-phase and grain-surface abundances of HCOOCH$_3$ and other molecules, for the shorter and longer times, respectively.  With the shorter time to reach a temperature of 200 K, the gas-phase routes to HCOOCH$_3$ become dominant over the surface reactions, whilst for the longer timescale, the surface route dominates even more strongly, and also produces the most methyl formate of any of the models. 

The major gas-phase route that forms methyl formate, reaction (5), is strongly dependent on the H$_2$CO abundance. With the shorter timescale, $t_{f} = 5 \times 10^{4}$ yr, formaldehyde is abundant until the end of the model. This is not the case with the longer timescale, $t_{f} = 1 \times 10^{6}$ yr, and here the gas-phase route acts for only a short time in comparison to the length of time that the HCOOCH$_3$ resides on the grain surface, gradually being destroyed by cosmic ray-induced photodissociation. However, the productivity of the grain-surface mechanism is {\em enhanced} by the longer timescale, because there is a longer period between the time when HCO becomes both abundant and highly mobile (i.e. following CO evaporation), and the time when formaldehyde evaporates -- the requirements for reaction (1) to be effective. Since the surface formaldehyde abundance is steady until it ultimately evaporates, and the OH required to form HCO (via OH + H$_2$CO $\rightarrow$ HCO + H$_2$O) is provided steadily by photodissociation of water ice, the amount of methyl formate formed by surface reactions is directly proportional to the timescale.

For the shorter timescale, the comparatively long lifetime of H$_2$CO in the gas phase means that most HCOOH is formed in the gas phase. Hence, shorter timescales favour the gas-phase routes, whereas longer timescales favour the grain-surface mechanism.

The final CH$_3$OCH$_3$ abundance suffers with the longer timescale, as photodissociation reduces the surface-based methanol abundance before it can evaporate and form dimethyl ether. However, the early peak in CH$_3$OCH$_3$ abundance caused by the evaporation of the surface species is almost as strong as the later peak induced by the gas-phase chemistry. Surface production is increased for similar reasons as stated for HCOOH and HCOOCH$_3$.

We also ran models using the linear temperature dependence, $\Delta T_1$, shown in figure \ref{figure2}, with the standard switch-on time. The difference between the two profiles is that for $\Delta T_2$, the increase in temperature is slower at early times, but then much faster at later times. Hence, with $\Delta T_1$, the hot core spends less time at lower temperatures and more time at higher temperatures. 
Because the surface reactions (1) -- (3) are disfavored for shorter timescales, and because these reactions are most active from $\sim$20 -- 60 K, the $\Delta T_1$ profile results in less surface formation since it increases more rapidly in this range of temperatures.

\section{Discussion}

Our results show that significant abundances of saturated organic molecules are produced in hot cores during the protostellar warm-up phase.  The results also demonstrate that the gas-phase and grain-surface chemistry during this period are strongly coupled. The large abundances which build up on grain surfaces have a large impact on gas-phase chemistry when they evaporate. But, also, the products of that chemistry can re-accrete, changing the surface chemistry. In addition, the sudden absence of certain species from the grains, following evaporation, can have a strong effect on the surface chemistry. Since hydrogenation is inefficient at warmer temperatures, the abundance of heavy radicals on the grain surfaces is mainly due to the photodissociation of ices such as H$_2$O, and to a lesser extent CH$_3$OH, due to cosmic ray-induced photons. The gas-phase and grain-surface chemistries act in different, complementary ways, and the strongly coupled system yields results which only a full gas-grain treatment can produce.

Table \ref{tab4} lists the gas-phase abundances obtained with various model parameters for methyl formate, formic acid, dimethyl ether, formaldehyde, and methanol at the time of maximum gaseous methyl formate abundance, which occurs at or near the end of the warm-up phase.  These abundances derive from both grain-surface and gas-phase reactions.  We have also listed the methyl formate abundances obtained without the gas-phase route, since reaction (5) is problematic.  If we look at the first three columns of the table, obtained with varying times for the warm-up phase but all with our standard  $\Delta T_{2}$ profile, we see that different species show different sensitivities to the timescale, which is related to the mass of the protostar.  Methyl formate increases one order of magnitude in abundance from the shortest to the longest timescale, with a stronger  dependence if the gas-phase synthesis is removed.  On the other hand, formic acid has an abundance that does not change much.  Dimethyl ether shows a rather strange dependence, first increasing then decreasing as the time interval increases.  The species with the strongest time dependence is formaldehyde, the abundance of which decreases by almost five orders of magnitude from the shortest to the longest time interval considered.  The abundance of methanol also decreases with increasing time interval, although not to as great an extent as that of formaldehyde.

\begin{table*}
\caption[]{Fractional abundances with respect to H$_2$, for some observed hot cores.$^{a}$}
\label{tab5}
\centering
\begin{tabular}{l c c c c c}
\hline\hline
\noalign{\smallskip}
Species              && Orion Compact Ridge$^{b}$ & Orion Hot Core$^{b}$ & IRAS 16293$^{c}$  & IRAS 4A$^{d}$ \\
\noalign{\smallskip}
\hline
\noalign{\smallskip}
HCOOCH$_3$-A         && 3$(-8)$                   & 1.4$(-8)$            & $1.7 \pm 0.7(-7)$ & $3.4 \pm 1.7(-8)$ \\
\noalign{\smallskip}             
HCOOCH$_3$-E         &&                           &                      & $2.3 \pm 0.8(-7)$ & $3.6 \pm 1.7(-8)$ \\
\noalign{\smallskip}
HCOOH                && 1.4$(-9)$                 & 8$(-10)$             & $\sim$$6.2(-8)$   & $4.6 \pm 7.9(-9)$ \\
\noalign{\smallskip} 
CH$_3$OCH$_3$        && 1.9$(-8)$                 & 8$(-9)$              & $2.4 \pm 3.7(-7)$ & $\leq$$2.8(-8)$ \\
\noalign{\smallskip} 
H$_2$CO              && 4$(-8)$                   & 7$(-9)$              & $6.0(-8)$$^{e}$   & $2(-8)$$^{f}$ \\
\noalign{\smallskip} 
CH$_3$OH             && 4$(-7)$                   & 1.4$(-7)$            & $3.0(-7)$$^{e}$   & $\leq$$7(-9)$ \\
\noalign{\smallskip}
\hline
\noalign{\smallskip}
\multicolumn{6}{l}{$^{a}$ $a(b) = a \times 10^{b}$} \\
\multicolumn{6}{l}{$^{b}$ Values from \cite{sutton95a}. Values are given for total HCOOCH$_3$ abundance.} \\
\multicolumn{6}{l}{$^{c}$ Values from \cite{cazaux03a}.} \\
\multicolumn{6}{l}{$^{d}$ Values from \cite{bottinelli04a}.} \\
\multicolumn{6}{l}{$^{e}$ Values from \cite{schoier02a}.} \\
\multicolumn{6}{l}{$^{f}$ Values from \cite{maret04a}.} \\
\end{tabular}
\end{table*}

Let us now consider the agreement between our new approach to hot core chemistry and observations of organic molecules in these sources.  Table \ref{tab5} shows observational results for two hot cores and two hot corinos -- IRAS 16293 and IRAS 4A  -- associated with low-mass protostars rather than the high-mass protostars typical of hot cores \cite[]{bottinelli04a}.  One can see immediately that observed abundances  for IRAS 16293 are larger than for the other cores.  However, there is some disagreement over the H$_2$ column densities associated with this source \cite[see e.g.][]{peeters06a}, and this may help to explain the discrepancy. Indeed, the sizes and morphologies of the emitting regions of hot cores are not well defined, and therefore observationally determined abundances may be strongly dependent on beam size.

For the Orion Compact Ridge and Hot Core, the observed abundances are best fit by a heat-up interval of between $2 \times 10^{5}$ yr and $1 \times 10^{6}$ yr.  Note, though, that the abundances listed in table \ref{tab4} refer to times somewhat less than the full interval.  The worst disagreement occurs with formic acid, where the calculated abundance is two orders of magnitude too high. One explanation for this discrepancy lies in the gas-phase neutral-neutral reaction (OH + H$_{2}$CO) leading to HCOOH; there is only a small amount of evidence for this channel (see, e.g., http://kinetics.nist.gov/index.php). Methanol is best fit at time intervals near the upper value while the calculated abundances of formaldehyde and dimethyl ether are too low at this time, and methyl formate is not very sensitive to changes of time in the interval range considered.  The hot corino IRAS 4A is somewhat richer in methyl formate and formic acid than the hot cores, but has less methanol.  Nevertheless, the level of agreement is not very different in the range of heat-up intervals considered.  For the case of IRAS 16293, the situation is rather unique.  Here the formic acid is in agreement with our large calculated abundance but the methyl formate abundance is more than an order of magnitude higher than our largest value and the dimethyl ether abundance is similarly large.  The formaldehyde and methanol abundances are not exceptionally large, and so rule out a very small heat-up interval.  In other words, both the hot cores and hot corinos in our sample are fit best by heat-up intervals characteristic of low-mass and intermediate-mass stars, although we cannot distinguish between the two.

Our results, while not in excellent agreement with observation and while only focusing on abundances of selected species, add strength to the case that hot cores develop over long timescales, spending significant periods at low to intermediate temperatures before reaching the $\sim$100 -- 300 K typical of observed objects.  Whilst our suggested heat-up timescales of $2 \times 10^{5}$ -- $1 \times 10^{6}$ yr are quite long, our model suggests that observed hot cores could have comparatively short post-evaporation ages, where we define the ``evaporation stage'' as that at which CH$_3$OH, HCOOH, HCOOCH$_3$, H$_2$O and NH$_3$ evaporate, approximately at the same time (at around 100 K). Note that this post-evaporation era is, according to previous models, when gas-phase chemistry is supposed to produce the complex species considered here \cite[see e.g.][]{millar91b,charnley92b,caselli93a}. 

Our standard and longer timescale models, $t_f = 2 \times 10^{5}$ and $t_f = 1 \times 10^{6}$ yr, spend approximately $3 \times 10^{4}$ and $3 \times 10^{5}$ yr, respectively, between the evaporation of the large molecules and the end of the model. This corresponds to a change in temperature from $\sim$100 K up to 200 K. The models largely produce acceptable agreement with observed abundances throughout this period, for the species which we investigate here, although for the longer timescale model methyl formate abundance has begun to depreciate seriously after around $10^{5}$ yr post-evaporation. The standard model displays acceptable HCOOCH$_3$ levels right to the end of the run, corresponding to $\sim$$3 \times 10^{4}$ yr post-evaporation.
Our results (see figures \ref{figure3} and \ref{figure5}) therefore suggest that hot cores which show HCOOCH$_3$ abundances of around 10$^{-8}$ with respect to H$_2$ could have post-evaporation ages of anywhere between zero and at least $3 \times 10^{4}$ yr (in the case of cores associated with intermediate-mass protostellar sources), and $1 - 2 \times 10^{5}$ yr (in the case of cores associated with low-mass protostellar sources).

Our model suggests that grain-surface formation of HCOOCH$_3$ occurs within a temperature range of $\sim 20$ -- $40$ K. The lower bound in this range corresponds to the evaporation of CO, which leaves the OH radical free to react with formaldehyde to form HCO. The upper bound corresponds to the evaporation of H$_2$CO, which is crucial to the formation of both of the radicals required by reaction (1). Hence, the binding energies of these two species are especially important in determining the time period available for reaction (1) to take place. The binding energy of HCO is much less important, since it is too reactive to build up on the grains; its abundance is determined by its rate of reaction with other species. Although the strength of its diffusion barrier is related to the binding energy ($E_{D} = 0.5 E_{B}$), we do not expect the {\em final} amount of HCOOCH$_3$ produced to be exceptionally sensitive to this value.

Following the experimental evidence of \cite{collings04a} for other species, \cite{viti04a} suggested that the evaporation of H$_2$CO is determined by co-desorption with water ice. In this model we allow thermal evaporation of H$_2$CO according to its desorption energy, less than half that of water. However, our model indicates that most formaldehyde should be formed at very late times in the collapse phase, so that H$_2$CO is unlikely to be trapped in the water ices below. Using our model to explicitly track the deposition of monolayers would allow us to investigate this more fully.

\section{Conclusions}

Using a gas-grain network of reactions, we have investigated the formation of methyl formate in hot cores, along with the related species dimethyl ether and formic acid, and to a lesser extent formaldehyde and methanol, during the protostellar warm-up and evaporative phase.  We undertook this work because standard models of hot-core chemistry, which rely on gas-phase chemistry at a temperature of 100 K or so to produce complex molecules from precursor methanol, have severe failings \cite[see e.g.][]{horn04a}. The chemistry during the warm-up period occurs both on grain surfaces and in the gas phase. We find that the formation of HCOOCH$_3$, by either mechanism, requires a gradual warm-up of temperatures, rather than an immediate jump from 10 K to 100 K, as is often assumed. In the case of the grain-surface route, reaction (1), this is so that radicals may become mobile enough to react, without quickly desorbing. Indeed, whilst the time-dependent temperature profiles we use are quite specific, grain-surface formation of methyl formate should be viable as long as temperatures remain in the 20 -- 40 K range for a long enough period of time. In the case of the only viable gas-phase route, reaction (5), it is because the reaction rate is strongly temperature dependent -- being fastest at low temperatures -- but requiring the evaporation of H$_2$CO at intermediate temperatures to become important. The plausibility of the gas-phase route for methyl formate identified in this work is a matter for further study.

During the warm-up phase, atomic hydrogen evaporates very quickly and hydrogenation is much less important than during the cold phase. The production of surface radicals is largely determined by cosmic ray-induced photodissociation of ices. However, in some cases, the low surface abundance of atomic H is still sufficient to produce radicals, when the other reactant is very abundant, e.g. formaldehyde.
 
Our results for a variety of species show that both gas-phase and grain-surface reactions are strongly coupled. Such a strong interaction allows processes to take place that could not happen on either the grain surfaces or in the gas phase alone.  
The results of our calculations show that towards the end of the warm-up period, large abundances of a variety of organic molecules are now present in the gas, in agreement with observation.  The observed rich, complex chemistries of hot cores may thus be representative of objects in which complex molecules have very recently evaporated (at around 100 K), contrary to the predictions of models that utilise only a gas-phase chemistry to reproduce all abundances \cite[e.g.][]{rodgers01a}.

The physical model of the warm-up phase, which we base on the analysis of Viti et al. (2004, and references therein), assumes a finite time period for the central protostellar heat source to reach its switch-on time, and therefore for the surrounding gas and dust to vary in temperature gradually. This means that the temperature increases should propagate outwards, and that the chemistries of the inner regions should be further advanced than the outer. Therefore, the typical hot core chemistry itself should propagate outwards. The radial size of regions which exhibit significant complex molecule abundances may then indicate the speed at which the temperature gradient propagates. 

We hope soon to extend our model to include a much larger network of surface reactions and species. This set will include other astronomically detected molecules such as the isomers of methyl formate and dimethyl ether, as well as some postulated and (so far) undetected species. Surface reaction of heavy radicals will also be extended to allow the formation of other large molecules already present in the gas-grain code.

\begin{acknowledgements}
We thank the National Science Foundation (US) for support of the Ohio State University astrochemistry program. Helpful discussions with S. Viti, on the subject of the warm-up phase, with M. Collings, on the subject of thermal desorption of grain mantles, with V. Wakelam, on the subject of atomic abundances, and with S. Widicus Weaver, on the subject of complex molecule formation, are acknowledged.
\end{acknowledgements}


\bibliographystyle{aa}
\bibliography{robbib_2006}

\end{document}